\documentclass[10pt,twocolumn,letterpaper]{article}

\usepackage{cvpr}              
\usepackage{multirow}
\definecolor{cvprblue}{rgb}{0.21,0.49,0.74}
\usepackage[pagebackref,breaklinks,colorlinks,allcolors=cvprblue]{hyperref}
\definecolor{myblue}{cmyk}{0,0.22,0.33,0}
\definecolor{mygreen}{cmyk}{0.76,0,0.76,0.30} 
\definecolor{myred}{cmyk}{0,1,1,0} 
\definecolor{lightorange}{RGB}{255,230,200} 

\usepackage{pifont}
\usepackage{booktabs}
\usepackage{colortbl}
\usepackage{xcolor}
\usepackage{algorithm}
\usepackage{algpseudocode}
\usepackage[accsupp]{axessibility}
\newcommand{\cmark}{\textcolor{mygreen}{\ding{51}}} 
\newcommand{\xmark}{\textcolor{myred}{\ding{55}}} 

\title{DRAMA: Next-Gen Dynamic Orchestration for Resilient Multi-Agent Ecosystems in Flux}

\author{
Xinkui Zhao$^{1,2,3}$,
Yifan Zhang$^{1}$,
Sai Liu$^{1}$,
Naibo Wang$^{2,1,3}$\textsuperscript{\thanks{Corresponding author.}},
Guanjie Cheng$^{2,1,3}$,
Yueshen Xu$^{4}$,\\
Chang Liu$^{2,1,3}$,
Shuiguang Deng$^{1}$,
Jianwei Yin$^{1}$\\ 
\small $^{1}$Zhejiang University, $^{2}$Ningbo Global Innovation Center, Zhejiang University \\
\small $^{3}$Zhejiang Key Laboratory of Digital-Intelligence Service Technology, $^{4}$Xidian University \\
\tt \small \{zhaoxinkui, 12451018, liusai2024, wangnaibo, chengguanjie\}@zju.edu.cn, ysxu@xidian.edu.cn,\\ 
\tt \small \{chang.liu, dengsg\}@zju.edu.cn, zjuyjw@cs.zju.edu.cn
}

\begin{document}
\maketitle
\begin{abstract}
Embodied Multi-Agent Systems (EMAS) have proven highly effective in addressing complex tasks through coordinated collaboration among heterogeneous agents. However, real-world environments and task specifications are inherently dynamic, exhibiting frequent changes, uncertainty, and variability. Despite these characteristics, most existing EMAS frameworks employ static architectures with fixed agent capabilities and rigid task allocation strategies, which substantially constrain their adaptability to evolving conditions. This inflexibility presents significant challenges to maintaining robust and efficient multi-agent cooperation in dynamic and unpredictable settings. 

To address these limitations, we propose \textbf{DRAMA}---Next-Gen \textbf{D}ynamic O\textbf{R}chestr\textbf{A}tion for Resilient \textbf{M}ulti-\textbf{A}gent Ecosystems tailored for rapidly changing environments. DRAMA adopts a multilayer architecture that incorporates three principal mechanisms: (\emph{i}) adaptive scheduling via an affinity-driven mechanism, (\emph{ii}) fault-tolerant continuity through hierarchical trust-chain task takeover, and (\emph{iii}) collective spatial intelligence that consolidates distributed observations for predictive reasoning. Together, these components enable event-triggered rescheduling and decentralized fault recovery, ensuring uninterrupted task execution amid agent arrivals, dropouts, or recoveries. Extensive experiments in the embodied VirtualHome-Social environment demonstrate that DRAMA achieves a 7\% improvement in runtime efficiency, and a 10\% increase in throughput compared with state-of-the-art baselines, while maintaining superior stability and robustness under dynamic agent populations.

\end{abstract}    
\section{Introduction}
\label{sec:intro}
The rapid advancement of Large Language Models (LLMs) has significantly broadened their capabilities across diverse domains, including natural language processing~\cite{achiam2023gpt, touvron2023llama, liu2024deepseek} and code generation~\cite{guo2024deepseek, li2023starcoder, zhu2024deepseek, chowdhery2023palm, fried2022incoder}. These advancements have concurrently accelerated progress in multi-agent systems (MAS), endowing agents with enhanced planning, reasoning, and autonomous collaboration abilities applicable to a variety of scenarios~\cite{rashid2020monotonic, hu2019simplified, guo2024large}. LLM-based multi-agent systems (LLM-MAS) have demonstrated outstanding performance in collaborative planning and emergent problem-solving across diverse tasks~\cite{qian2023chatdev, hong2023metagpt, chen2024scalable}. When integrated into physical or simulated environments, such intelligent agents become embodied, enabling them to address long-horizon, open-world tasks that increasingly mirror human-like behavioral patterns~\cite{park2023generative, kaiya2023lyfe, cai2024digital, qin2024mp5}.

Despite these advances, most existing Embodied Multi-Agent System (EMAS) frameworks are built upon static architectures in which agent capabilities and task allocations remain fixed. This rigidity severely constrains adaptability and robustness in dynamic, real-world environments. In practice, both team composition and task conditions are inherently fluid—agents may join, depart, or fail, and task specifications can evolve unpredictably~\cite{dias2004robust, zhou2023robust}. Such volatility underscores the necessity of EMAS frameworks that support adaptive and resilient task allocation, thereby maintaining effective collaboration and stable performance amid continual changes in team structure and task demands. Addressing these challenges is crucial for the practical deployment of MAS in complex and evolving environments.

To address these limitations, we propose \textbf{DRAMA}—Next-Gen \textbf{D}ynamic O\textbf{R}chestr\textbf{A}tion for Resilient \textbf{M}ulti-\textbf{A}gent Ecosystems in Flux. DRAMA is built upon a multilayer architecture that integrates adaptive strategic layer, collective intelligence layer, and autonomous layer. In the \textbf{Strategic Layer}, an affinity matrix models real-time agent–task compatibility and is optimized using the Hungarian algorithm to achieve globally coherent task allocation. A hierarchical \textbf{trust-chain} mechanism ensures fault-tolerant continuity, where each task is protected by an ordered set of backup agents capable of immediate takeover in case of failure. The \textbf{Collective Intelligence Layer} aggregates spatial observations from distributed agents to learn probabilistic object–location priors, thereby enhancing perception and planning efficiency. Finally, the \textbf{Autonomous Layer} equips each agent with dual roles as executor and guardian, enabling real-time perception, action execution, and peer monitoring.

We evaluate DRAMA extensively in the embodied environment VirtualHome-Social~\cite{puig2018virtualhome}, benchmarking it against representative EMAS baselines. Results demonstrate that DRAMA is the only framework capable of reliably handling both agent dropout and recovery scenarios. Moreover, it achieves a 17\% improvement in runtime efficiency, a 13\% reduction in task conflict, and a 10\% gain in throughput compared to existing methods, while maintaining superior robustness and adaptability under frequent agent turnover and dynamic task conditions.

The main contributions are summarized as follows:
\begin{itemize}
    \item We propose \textbf{DRAMA}, the Next-Gen \textbf{D}ynamic O\textbf{R}chestr\textbf{A}tion for Resilient \textbf{M}ulti-\textbf{A}gent Ecosystems in Flux that supports real-time adaptation to agent arrivals, dropouts, and recoveries through event-triggered task reallocation and flexible resource management.
    
    \item We introduce a unified abstraction that models both agents and tasks as resource entities, enabling affinity-driven scheduling via the Hungarian algorithm and promoting modular, decoupled coordination in heterogeneous multi-agent environments.
    
    \item We design a three-layer architecture that integrates adaptive scheduling, collective intelligence, and fault-tolerant execution through trust-chain–based task takeover and decentralized collaboration.
    
    \item We conduct extensive experiments in the embodied VirtualHome-Social environment, demonstrating that DRAMA consistently outperforms state-of-the-art baselines in efficiency, resource utilization, and robustness.
\end{itemize}
\section{Background}
\label{sec:formatting}

\subsection{Embodied Multi-Agent Systems}

An EMAS consists of a collection of autonomous agents $\mathcal{A} = \{a_1, a_2, \dots, a_N\}$ that cooperatively or competitively perform a set of tasks $\mathcal{Q} = \{\mathtt{q}_1, \mathtt{q}_2, \dots, \mathtt{q}_M\}$ in a shared environment $\mathcal{E}$. Each embodied agent $a_i$  functions as an autonomous control loop interacting with the environment and can be formally represented as \(a_i = \langle S_i, O_i, A_i, \pi_i \rangle\), 
where $S_i$ denotes its local state space, $O_i$ its observation space, $A_i$ its available action set, and $\pi_i$ its policy for decision-making and control.  
The allocation of agents to tasks at time $t$ can be described by a mapping $f_t : \mathcal{Q} \rightarrow \mathcal{A}$, where $f_t(\mathtt{q}_j)$ specifies the agent responsible for task $\mathtt{q}_j$.  
The global system objective over time horizon $T$ is to maximize a collective utility function:
\begin{equation}
    \max_{\{f_t\}_{t=1}^T,\,\{\pi_i\}_{i=1}^N} 
    U(\{f_t\}_{t=1}^T,\mathcal{A},\mathcal{Q},\mathcal{E},T),
\end{equation}
where $U(\cdot)$ evaluates the overall task performance.

\begin{figure*}
    \centering
    \includegraphics[width=\linewidth]{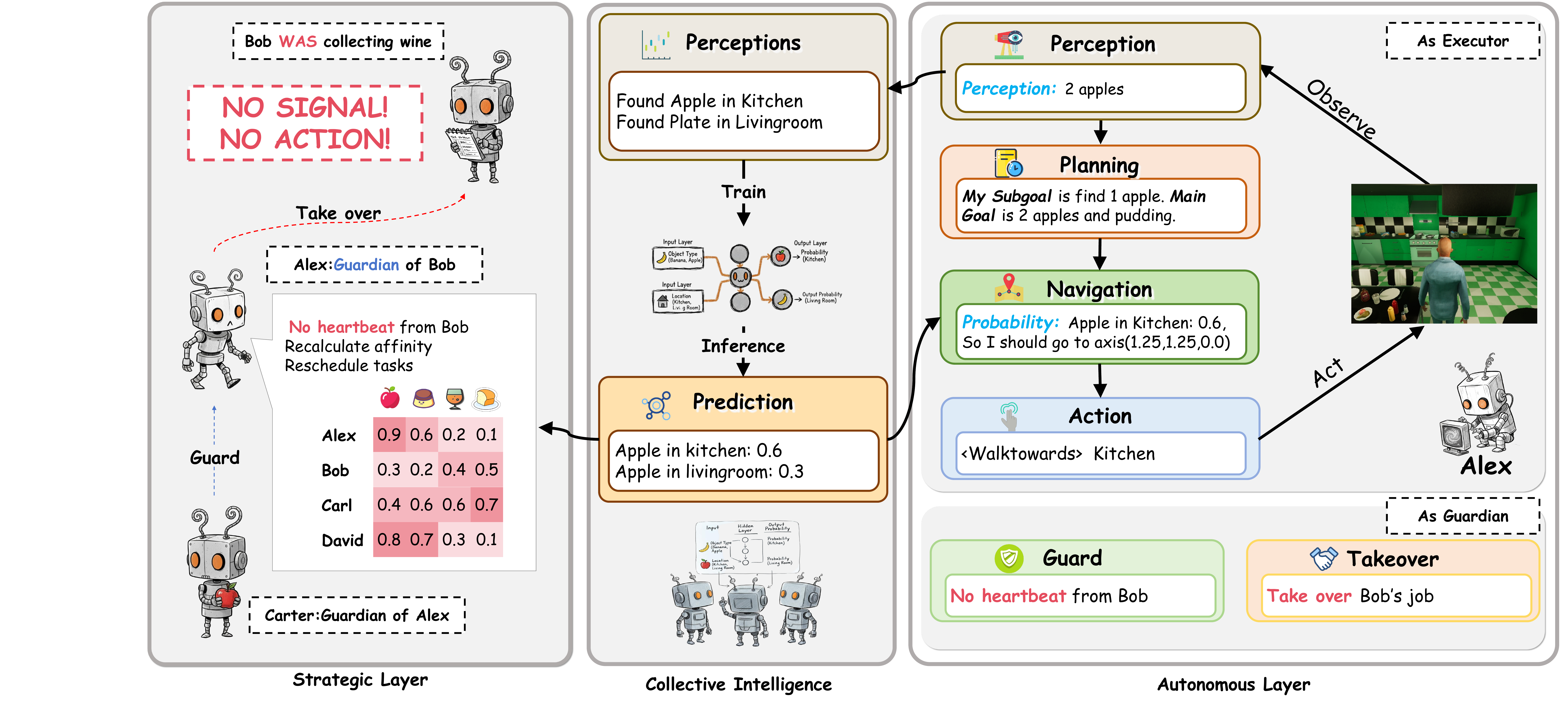}
    \caption{\textbf{Overview of the DRAMA framework}. It adopts a three-layer architecture—Strategic, Collective Intelligence, and Autonomous layers—for adaptive scheduling, collective spatial reasoning, and resilient task execution under dynamic agent arrivals and dropouts.}
    \label{fig:overview2}
\end{figure*}
\subsection{Motivation}

Although the general formulation of EMAS supports dynamic agent-task assignments over time, most existing frameworks adopt a static assignment paradigm in practice. Specifically, the assignment function is determined at initialization ($t = 0$) and remains fixed throughout the system's operation, i.e.,

\begin{equation}
    f_t = f_0, \quad \forall t \in [1, T].
\end{equation}
The system objective thus reduces to:
\begin{equation}
    \max_{f_0,\,\{\pi_i\}_{i=1}^N} \;\; U(f_0,\, \mathcal{A},\, \mathcal{Q},\, \mathcal{E},\, T)
\end{equation}

While the static assignment approach simplifies system design and deployment, it introduces several critical limitations in dynamic, real-world environments:

\begin{itemize}
    
    \item \textbf{Agent and Task Dynamism:} Agents may join, leave, or change capabilities over time; tasks may arrive, evolve, or be reprioritized. Static assignments prevent the system from adapting to these changes.

    \item \textbf{Suboptimal Resource Utilization:} Fixed assignments can result in idle agents, overloaded tasks, or inefficient execution, particularly as environmental conditions shift.

    \item \textbf{Lack of Real-Time Responsiveness:} Without mechanisms to monitor and adapt $f_t$, the system cannot respond effectively to failures, delays, or unforeseen events.

    \item \textbf{Limited Scalability and Robustness:} The inability to reallocate resources dynamically constrains system performance and resilience in open or uncertain settings.
\end{itemize}

These challenges underscore the necessity for EMAS architectures that enable adaptive, real-time agent-task assignment, leveraging continuous monitoring and feedback to optimize system utility in dynamic environments.

\section{Methodology}

In this section, we present \textbf{DRAMA}, a modular \textbf{three-layer framework} that unifies agents and tasks as resource entities for adaptive and resilient multi-agent orchestration. The architecture comprises: (i) a \textit{Strategic Layer} for affinity-based task allocation and fault-tolerant scheduling; (ii) a \textit{Collective Intelligence Layer} that aggregates distributed perceptions for shared spatial reasoning; and (iii) an \textit{Autonomous Layer} enabling agents to execute tasks and monitor peers. Together, these layers support real-time adaptation and continuous task execution under dynamic team compositions and environments. An overview is illustrated in Figure~\ref{fig:overview2}.

\subsection{Strategic Layer}
\label{sec:sl}
\begin{figure*}
    \centering
    \includegraphics[width=1\linewidth]{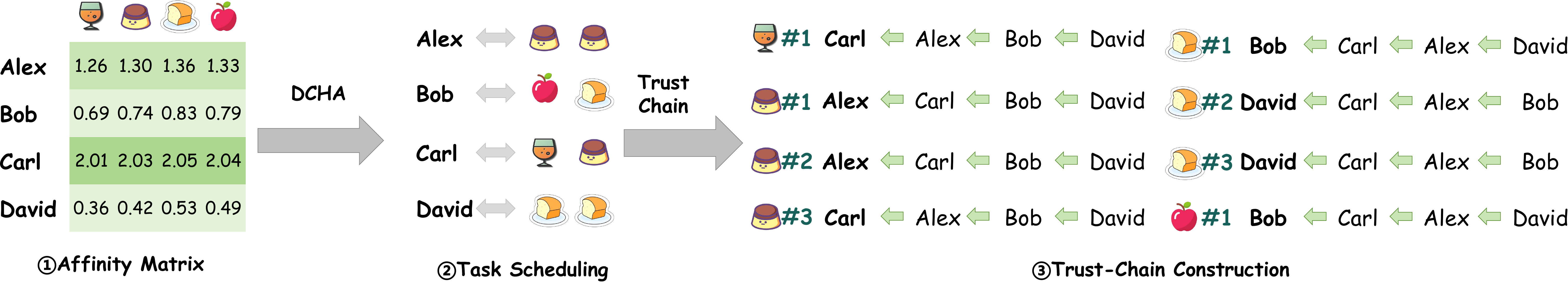}
    \caption{
        \textbf{A demonstration of the task scheduling and trust-chain mechanism in the Strategic Layer.}
        The left matrix shows dynamic affinity scores between agents and tasks, derived from historical cooperation and current context.
        The DCHA then produces optimal one-to-many task assignments (middle).
        Each assigned agent becomes the head of a hierarchical \textit{trust chain} (right), 
        where subsequent agents act as guardians to monitor and take over if predecessors fail.
        This design enables adaptive and fault-tolerant multi-agent scheduling.}
        \label{fig:strategic}
\end{figure*}

Building upon the multi-agent formulation introduced in Section~\ref{sec:formatting}, 
the \textbf{strategic layer} governs the adaptive optimization of the agent--task mapping $f_t$ 
and reinforces robustness through decentralized fault tolerance. 
As depicted in Figure~\ref{fig:strategic}, 
this layer integrates three components: 
(1) an affinity matrix capturing agent--task compatibility, 
(2) task scheduling via the Dual-Capacity Hungarian Assignment algorithm, and 
(3) trust-chain construction that provides hierarchical recovery for resilient task execution.
\paragraph{Affinity Computation.}
At time step~$t$, the instantaneous affinity between agent~$a_i$ and task~$q_j$ is formulated as a weighted combination of the agent’s available workload capacity and the predicted spatial distance between them:
\begin{equation}
\mathcal{S}_{ij,t}
= w_1 \cdot v_i(t)
- w_2 \cdot \mathrm{Dist}\big(\mathbf{x}_i(t), \hat{\mathbf{p}}_j(t)\big),
\end{equation}
where $v_i(t)$ represents the current availability of agent~$a_i$,
$\mathbf{x}_i(t)$ denotes its current position, and
$\hat{\mathbf{p}}_j(t)$ indicates the predicted location of task~$q_j$, which is estimated by the collective intelligence layer based on the spatial distribution~$P(o_j, p)$.
$\mathrm{Dist}(\cdot)$ refers to the Euclidean distance, and $w_1, w_2 > 0$ are weighting coefficients (see Section~\ref{sec:cil}).

\begin{algorithm}[t!]
\caption{Dual-Capacity Hungarian Assignment }
\label{alg:dcha}
\small
\begin{algorithmic}[1]
\Require Agents $A = \{a_i\}_{i=1}^N$, Tasks $T = \{t_j\}_{j=1}^M$, Affinity function $\text{Aff}$
\Ensure Mapping $\mathcal{M}: A \to \mathcal{P}(T)$ where $|\mathcal{M}(a_i)| \le 2$

\State $V \gets \emptyset$
\For{each agent $a_i \in A$}
    \State $V \gets V \cup \{v_{i,1}, v_{i,2}\}$ \Comment{Each agent provides two slots}
\EndFor

\Statex \textbf{1. Create virtual agents and cost matrix}
\State $C \gets$ new $|V|\!\times\!|T|$ cost matrix
\For{each virtual slot $v_{i,k} \in V$}
    \For{each task $t_j \in T$}
        \State $C[v_{i,k}, t_j] \gets -\text{Aff}(a_i, t_j)$
    \EndFor
\EndFor

\Statex \textbf{2. Solve as a standard assignment problem}
\State $K \gets \max(|V|, |T|)$
\State $C' \gets$ new $K\!\times\!K$ matrix, filled with $\infty$
\State $C'_{1..|V|,1..|T|} \gets C$
\State $P \gets \text{HungarianAlg}(C')$ \Comment{$P$ is a set of optimal pairs}
\Statex \textbf{3. Consolidate results back to real agents}
\State $\mathcal{M}(a_i) \gets \emptyset$ for all $a_i \in A$
\For{each pair $(v_{i,k}, t_j) \in P$}
    \If{$t_j \in T$} \Comment{Filter dummy tasks}
        \State $\mathcal{M}(a_i) \gets \mathcal{M}(a_i) \cup \{t_j\}$
    \EndIf
\EndFor

\State \Return $\mathcal{M}$
\end{algorithmic}
\end{algorithm}

\paragraph{Task Scheduling}
To achieve globally coherent task allocation, the system searches for the assignment mapping
that maximizes the total affinity over all tasks:
\begin{equation}
    f_t^* = 
    \arg\max_{f_t}
    \sum_{j=1}^{M} 
    \mathcal{S}_{f_t(q_j)j,t},
    \quad 
    \text{s.t. } f_t: \mathcal{Q} \to \mathcal{A} \text{ (one-to-one)}.
\end{equation}
The optimal mapping $f_t^*$ is determined by our proposed Dual-Capacity Hungarian Assignment (DCHA) algorithm. This method adapts the classic Hungarian algorithm~\cite{mills2007dynamic} for our one-to-many assignment problem, ensuring a globally optimal allocation that maximizes the total context-aware suitability. The full procedure is detailed in Algorithm~\ref{alg:dcha}.

\paragraph{Trust-Chain Construction.}
After the primary task assignment is obtained from the DCHA algorithm, 
the system constructs a hierarchical trust chain to provide fault-tolerant continuity for each task. 
For task $q_j$, let $\mathcal{S}_{:,j,t}$ denote the affinity column corresponding to all agents. 
Agents are ranked in descending order of affinity:
\[
a_{(1)} \succ a_{(2)} \succ a_{(3)} \succ \cdots \succ a_{(N)}.
\]
The agent selected by the DCHA algorithm,
$
a_{(1)} = f_t^*(q_j),
$
serves as the \textit{primary executor} and occupies the head of the trust chain.
Subsequent agents are organized into an ordered guardian sequence:
\[
a_{(1)} \rightarrow a_{(2)} \rightarrow a_{(3)} \rightarrow \cdots \rightarrow a_{(N)},
\]
where the arrow denotes the guarding direction.
Each guardian supervises the operational state of its predecessor
and remains on standby for takeover.

If the executor $a_{(1)}$ fails, 
responsibility is immediately transferred to its guardian $a_{(2)}$;
if $a_{(2)}$ fails subsequently, $a_{(3)}$ becomes active, and so forth.
This design ensures that the executor is always positioned as the foremost node in its guardian chain—the entity being protected—
and that every task possesses an ordered backup sequence for recovery.  
After each takeover event, local recalibration updates workload balance and affinity relationships to maintain coherent system scheduling.

\subsection{Collective Intelligence Layer}
\label{sec:cil}
The \textbf{collective intelligence layer} functions as the shared spatial cognition component of the system.
It integrates distributed observations from multiple agents,
learns spatial regularities of object occurrences across environments,
and provides predictive spatial priors of task or object locations
to support the strategic layer.

\paragraph{Collective Experience Integration}
Each agent~$a_i$ initially explores the environment to accumulate spatial experience through \textit{visual observations}, represented as  $\mathcal{H}_{i,t} = \{(o_k, p_k) \mid k < t \}$, where $o_k$ denotes a visually perceived object or task type, and $p_k$ indicates its corresponding spatial coordinates (e.g., room or location). After several explorations or task execution, all agents exchange and integrate their visual observation histories as $\mathcal{H}_t = \bigcup_{i=1}^{N} \mathcal{H}_{i,t}$. This shared dataset constitutes a collective spatial memory that captures the co-occurrence relationships between objects and locations, serving as the empirical foundation for estimating spatial distributions $\mathcal{P}(o,p)$.

\paragraph{Collective Spatial Inference}
Based on the aggregated observation history~$\mathcal{H}_t$,
the collective intelligence layer learns the spatial likelihood of objects or tasks across the environment.
Rather than relying solely on empirical frequency estimation, 
which provides only a coarse prior, we train an MLP-based predictor $f_\theta(\cdot)$ 
to approximate the joint distribution $\mathcal{P}(o,p)$ capturing higher-order spatial correlations between contextual factors and object locations. 
Given contextual input $x_t$ (e.g., current room features, semantic cues, and co-occurring objects),
the network outputs the probability of an object~$o$ appearing at position~$p$:
\[
\hat{\mathcal{P}}_\theta(o,p \mid x_t) = f_\theta(x_t).
\]
The predicted most probable location of a queried object or task~$o_j$ is then obtained as
$\hat{p}_j = \mathbb{E}_{p\sim \hat{\mathcal{P}}_\theta(o_j,p)}[p]$ or $\arg\max_p \hat{\mathcal{P}}_\theta(o_j,p)$.
This design enables generalization beyond explicitly observed samples 
and supports inference in unexplored areas. See details in the supplementary.




\subsection{Autonomous Layer}
The \textbf{autonomous layer} constitutes the self-regulated control architecture of each agent. 
It integrates perceptual processing, goal planning, motion execution, and fault monitoring 
to achieve adaptive and resilient autonomy. 
Every agent functions simultaneously as an \emph{executor}, 
responsible for perception and action, and as a \emph{guardian}, 
responsible for peer supervision and fault handling.

\paragraph{Executor Role}
In the executor pathway, the autonomous layer establishes a continuous perception–action loop that grounds high-level goals in concrete environmental interactions. 
Each agent maintains a \emph{temporal context} by systematically recording its past actions, observations, communications, and the evolving state of its surroundings. 
This persistent memory allows agents to adapt their behavior based on accumulated experience. 
To manage increasing task complexity and the expanding history of observations, agents employ a \emph{hierarchical memory management} strategy: 
information related to ongoing primary and critical subtasks is retained in detail to support timely decision-making, 
recent but less crucial events are compressed into summarized representations, 
and outdated or irrelevant data are selectively discarded to conserve computational resources.

Upon receiving a coarse-grained task from the control plane, the agent decomposes the high-level objective into finer-grained, actionable subtasks tailored to the current environment.
This process leverages real-time perceptual inputs—such as spatial layout, object locations, and environmental changes—together with memory of past actions and outcomes. 
The \emph{perception module} transforms raw sensory information into structured scene representations for reasoning, 
the \emph{planning module} formulates executable subgoals, 
and the \emph{navigation module} integrates predicted object or task locations $\hat{p}_j$ from the collective intelligence layer to generate trajectory plans and motion commands. 
Finally, the \emph{action module} executes these plans through motor control, directly interacting with the environment.

\paragraph{Guardian Role}
Running parallel to the executor process, the guardian mechanism safeguards operational robustness and continuity. 
The \emph{guard module} continuously monitors heartbeat signals and status indicators from peer agents to detect anomalies such as latency, dropouts, or malfunction. 
When such irregularities are identified, 
the \emph{takeover module} initiates a decentralized recovery procedure: it evaluates the pending responsibilities of the affected agent, reallocates unfinished tasks to suitable peers based on current workload and spatial proximity, and triggers re-execution through their executor pathways. 
This trust-chain–driven reallocation ensures seamless task recovery, maintains system integrity, and preserves global performance under agent failures or dynamic team composition.
\section{Experiment Setup}

\subsection{Dataset}
To comprehensively evaluate the effectiveness and robustness of various embodied multi-agent coordination methods, we build upon the VirtualHome-Social environment~\cite{puig2018virtualhome, zhang2023building, puig2020watch}, a Unity-based simulation platform. We observe that the default task configurations in C-WAH~\cite{puig2020watch} are relatively simple and may not adequately distinguish the capabilities of advanced approaches. To address this limitation, we design a more challenging experimental setting by increasing both the number of objects and the diversity of task types, thereby introducing greater complexity and requiring more sophisticated cooperation strategies. All trials are repeated in the event of simulator-induced failures. Details of the dataset are provided in the supplementary.

\subsection{Baselines}

To comprehensively evaluate our framework, we compare it against several representative baselines, each reflecting a distinct coordination or planning paradigm. Since these baselines were not originally designed for VirtualHome-Social environment, we adapted and re-implemented them for this setting. For robustness evaluation, we adhere to each baseline’s original design, including their lack of support for task handover where applicable. For efficiency evaluation, however, we equip all methods with additional capabilities to ensure comparably high success rates: specifically, each method is provided with both the final target and current progress information as input. These enhancements enable a fair comparison of the efficiency of different task assignment strategies across baselines.

The compared methods are:

\begin{itemize}
    \item \textbf{CoELA}~\cite{zhang2023building}: Enables agents to plan, communicate, and cooperate via inter-agent messaging and teamwork.
    \item \textbf{MCTS}~\cite{gan2025master}: Employs Monte Carlo Tree Search for planning-based task allocation, selecting high-reward solutions within a fixed search budget.
    \item \textbf{AgentVerse-Static (AV-Static)}~\cite{chen2023agentverse}: Assigns agent roles and tasks only at initialization; assignments remain fixed.
    \item \textbf{AgentVerse-Dynamic (AV-Dynamic)}: Triggers reallocation among agents only upon task completion, periodically redistributing remaining tasks.
    \item \textbf{ProAgent}~\cite{zhang2024proagent}: Adopts a fully decentralized approach—agents coordinate via communication, prediction, and negotiation, without a central scheduler.
\end{itemize}

\subsection{Metrics}
\begin{table*}[!ht]
\centering
\caption{\textbf{Effectiveness comparison across different scenarios and methods.} 
Each method is evaluated with three metrics: Average Steps (AS), Conflict Rate (CR), and Throughput (TP).}
\label{tab:merged_performance}
\resizebox{\textwidth}{!}{%
\begin{tabular}{lcccccccccc}
\toprule
& & \multicolumn{3}{c}{\textbf{Static}} & \multicolumn{2}{c}{\textbf{Dropout}} & \multicolumn{2}{c}{\textbf{Addition}} & \multicolumn{1}{c}{\textbf{Recovery}}& \multicolumn{1}{c}{\textbf{Replacement}} \\
\cmidrule(lr){3-5} \cmidrule(lr){6-7} \cmidrule(lr){8-9} \cmidrule(lr){10-10}  \cmidrule(lr){11-11}
\textbf{Method} & \textbf{Metrics} 
& \textbf{3} & \textbf{4} & \textbf{5}
& \textbf{4$\rightarrow$3} & \textbf{5$\rightarrow$4$\rightarrow$3}
& \textbf{4$\rightarrow$5} & \textbf{3$\rightarrow$4$\rightarrow$5}
& \textbf{4$\rightarrow$3$\rightarrow$4} & \textbf{4$\rightarrow$3$\rightarrow$4$^{\prime}$}\\

\rowcolor{gray!20}
\multicolumn{11}{c}{\textit{Comparison Methods}} \\

\multirow{3}{*}{CoELA}
& AS & 70.511 & 54.956 & 52.293 & 61.841 & 57.556 & 54.415 & 57.565 & 60.791 & 60.977 \\
& CR & 0.0614 & 0.0883 & 0.173 & 0.065 & 0.102 & 0.121 & 0.100 & 0.085 & 0.090 \\
& TP & \cellcolor{gray!10}0.169 & \cellcolor{gray!10}0.222 & 0.240 & 0.182 & 0.195 & 0.228 & 0.198 & 0.189 & 0.182 \\
\midrule

\multirow{3}{*}{AV-dynamic}
& AS & 72.784 & 56.296 & 47.974 & \cellcolor{gray!10}59.297 & 56.244 & 54.732 & 61.342 & 59.000 & 61.422 \\
& CR & 0.060 & 0.074 & 0.128 & 0.048 & 0.089 & 0.100 & 0.076 & 0.062 & 0.068 \\
& TP & 0.151 & 0.203 & \cellcolor{gray!10}0.241 & 0.193 & 0.204 & \cellcolor{gray!10}0.234 & \cellcolor{gray!10}0.222 & 0.198 & 0.193 \\
\midrule

\multirow{3}{*}{AV-static}
& AS & 68.447 & 52.727 & 49.450 & 60.568 & 55.024 & 51.539 & 56.846 & 56.822 & 59.781 \\
& CR & 0.071 & 0.095 & 0.122 & 0.058 & 0.108 & 0.106 & 0.084 & 0.064 & 0.073 \\
& TP & 0.160 & 0.212 & 0.240 & 0.195 & 0.207 & 0.230 & 0.221 & 0.207 & 0.190 \\
\midrule

\multirow{3}{*}{ProAgent}
& AS & \cellcolor{gray!10}64.861 & \cellcolor{gray!10}51.560 & \cellcolor{gray!10}47.683 & 59.761 & \cellcolor{gray!10}54.955 & \cellcolor{gray!10}48.717 & \cellcolor{gray!10}54.804 & \cellcolor{gray!10}55.296 & \cellcolor{gray!10}56.711 \\
& CR & 0.038 & 0.064 & 0.109 & 0.065 & 0.111 & 0.136 & 0.092 & 0.080 & 0.087 \\
& TP & 0.166 & 0.211 & 0.237 & \cellcolor{gray!10}0.196 & \cellcolor{gray!10}0.208 & 0.233 & 0.205 & \cellcolor{gray!10}0.214 & \cellcolor{gray!10}0.197 \\
\midrule

\multirow{3}{*}{MCTS}
& AS & 92.200 & 77.330 & 67.260 & 88.780 & 86.850 & 72.000 & 80.160 & 83.330 & 82.070 \\
& CR & \cellcolor{gray!10}0.033 & \cellcolor{gray!10}0.023 & \cellcolor{gray!10}0.044 & \cellcolor{gray!10}0.035 & \cellcolor{gray!10}0.026 & \cellcolor{gray!10}0.019 & \cellcolor{gray!10}0.014 & \cellcolor{gray!10}0.028 & \cellcolor{gray!10}0.029 \\
& TP & 0.133 & 0.158 & 0.174 & 0.138 & 0.105 & 0.151 & 0.111 & 0.113 & 0.114 \\

\rowcolor{gray!20}
\multicolumn{11}{c}{\textit{Our Method}} \\

\multirow{3}{*}{DRAMA}
& AS & \textbf{59.977} & \textbf{50.242} & \textbf{46.697} & \textbf{57.205} & \textbf{52.796} & \textbf{48.213} & \textbf{53.367} & \textbf{55.069} & \textbf{55.429} \\
& CR & \textbf{0.027} & \textbf{0.062} & \textbf{0.083} & \textbf{0.026} & \textbf{0.076} & \textbf{0.068} & \textbf{0.057} & \textbf{0.044} & \textbf{0.061} \\
& TP & \textbf{0.189} & \textbf{0.217} & \textbf{0.252} & \textbf{0.200} & \textbf{0.216} & \textbf{0.235} & \textbf{0.211} & \textbf{0.204} & \textbf{0.205} \\
\bottomrule
\end{tabular}%
}
\end{table*}

To rigorously assess multi-agent cooperation in dynamic environments, we employ a comprehensive set of metrics that evaluate both outcome and process-level performance. These metrics capture not only task completion but also system efficiency, adaptability, and resource utilization.






\textbf{Average Steps (AS):}
Quantifies temporal efficiency by calculating the mean number of steps taken by the agent completing the final task in each episode, indicating overall latency and the system’s \emph{effectiveness} in minimizing completion time for complex tasks.

\textbf{Conflict Rate (CR):} 
Evaluates the degree of coordination by measuring the proportion of steps in which two or more agents 
simultaneously execute the same action on the same object. 
A higher conflict rate indicates redundant or conflicting behaviors, while a lower rate reflects stronger cooperative synchronization.

\textbf{Throughput (TP):}
Measures task-level productivity by computing the ratio between the number of successfully completed target objects 
and the total number of steps. 
It reflects both speed and consistency in task execution.


\subsection{Scenario Design}

To thoroughly evaluate the robustness and adaptability of our framework, 
we design a set of experimental scenarios that cover both stable and dynamic agent populations. 
In our implementation, the maximum number of embodied agents is limited to five, 
representing the upper bound of the current system capacity.

\begin{itemize}
    \item \textbf{Static Scenarios:} 
    The agent population remains fixed throughout the experiment. 
    We consider three static configurations consisting of 3, 4, and 5 agents, 
    which respectively represent small, medium, and full-capacity settings.

    \item \textbf{Dynamic Scenarios:} 
    To evaluate the framework's adaptability, we designed scenarios simulating agent failures, resource scaling, and team turnover.
    
      The dynamic scenarios are categorized as follows:
    
    \begin{itemize}
    \item \emph{Agent Dropout:} Agents are removed with patterns of $4 \!\rightarrow\! 3$ and cascading $5 \!\rightarrow\! 4 \!\rightarrow\! 3$.
    
    \item \emph{Agent Addition:} New agents are introduced with patterns of $4 \!\rightarrow\! 5$ and $3 \!\rightarrow\! 4 \!\rightarrow\! 5$.
    
    \item \emph{Agent Turnover:} We test two complex turnover events:
    \begin{itemize}
        \item \emph{Recovery:} An agent temporarily leaves and later rejoins at same location($4 \!\rightarrow\! 3 \!\rightarrow\! 4$).
        
        \item \emph{Replacement:} An agent leaves and is replaced by a new agent at a different location ($4 \!\rightarrow\! 3 \!\rightarrow\! 4'$).
    \end{itemize}
\end{itemize}
\end{itemize}

To ensure fair comparison across experiments, 
the timing of agent removal or addition is aligned to fixed progress ratios of the overall task execution, so that all methods experience structural changes at equivalent temporal stages. These scenarios collectively evaluate the framework’s ability to maintain stability, 
adapt to resource fluctuations, and recover from agent interruptions under varying population scales.

\section{Experimental Results}

\subsection{Effectiveness of DRAMA}
\paragraph{Efficiency.}
To evaluate the efficiency of \textbf{DRAMA}, we compare the Average Steps (AS) across all methods, as presented in Table~\ref{tab:merged_performance}. In static settings, \textbf{DRAMA} consistently achieves the lowest AS, indicating clear improvements in execution efficiency. Under dynamic conditions—including agent dropout, addition, recovery, and replacement—\textbf{DRAMA} maintains robust superiority, reducing AS by approximately 4\% on average and achieving up to 7--8\% improvement in highly dynamic scenarios. These results demonstrate that \textbf{DRAMA} substantially enhances operational efficiency and responsiveness in evolving multi-agent environments.

\paragraph{Conflict Rate.}
Although the rule-based MCTS framework achieves slightly lower CR due to its static and pre-defined scheduling policy, it cannot adapt to real-time agent or task variations. Among all \textit{dynamic scheduling} frameworks, \textbf{DRAMA} consistently attains the lowest CR across both static and dynamic settings, demonstrating superior coordination and reduced inter-agent conflicts.

\paragraph{Throughput.}
Beyond efficiency and coordination, \textbf{DRAMA} also achieves the highest throughput across all scenarios, with an overall \textbf{10--20\%} improvement over existing methods. This gain indicates that DRAMA not only accelerates task completion but also sustains high productivity and consistent performance under fluctuating agent populations and dynamic task conditions.

\begin{table}[t]
\centering
\caption{
Scenario success comparison across different environmental conditions. 
\cmark~indicates success and \xmark~indicates failure. 
\textbf{DRAMA} successfully supports all scenario types.
}
\resizebox{\linewidth}{!}{
\renewcommand{\arraystretch}{1.1}
\setlength{\tabcolsep}{3pt}
\begin{tabular}{cccccc>{\columncolor{lightorange}}c}
\toprule
\textbf{Scenario} & \textbf{CoELA} & \textbf{MCTS} & \textbf{ProAgent} & \textbf{AV-static} & \textbf{AV-dynamic} & \textbf{DRAMA} \\
\midrule
Static & \cmark & \cmark & \cmark & \cmark & \cmark & \cmark \\
Dropout & \xmark & \xmark & \xmark & \xmark & \xmark & \cmark \\
Addition & \cmark & \cmark & \cmark & \cmark & \cmark & \cmark \\
Recovery & \xmark & \xmark & \xmark & \xmark & \xmark & \cmark \\
\bottomrule
\end{tabular}
}

\label{tab:dynamic_sr}
\end{table}
\subsection{Robustness of DRAMA}

To comprehensively illustrate framework-level robustness, Table~\ref{tab:dynamic_sr} summarizes whether each method successfully completes the task across a variety of environmental scenarios, including static settings and different types of agent dynamics. Specifically, the dropout scenarios simulate reductions in the number of agents from 5 to 3, while the addition scenario involves an increase from 3 to 5 agents. Notably, DRAMA is the only framework that successfully addresses all scenarios. In contrast, all baseline methods fail under agent dropout conditions, as they lack the capacity to reassign incomplete tasks when agents leave.

\subsection{Stableness of DRAMA}

\begin{figure}[t]
    \centering
    \includegraphics[width=1\linewidth]{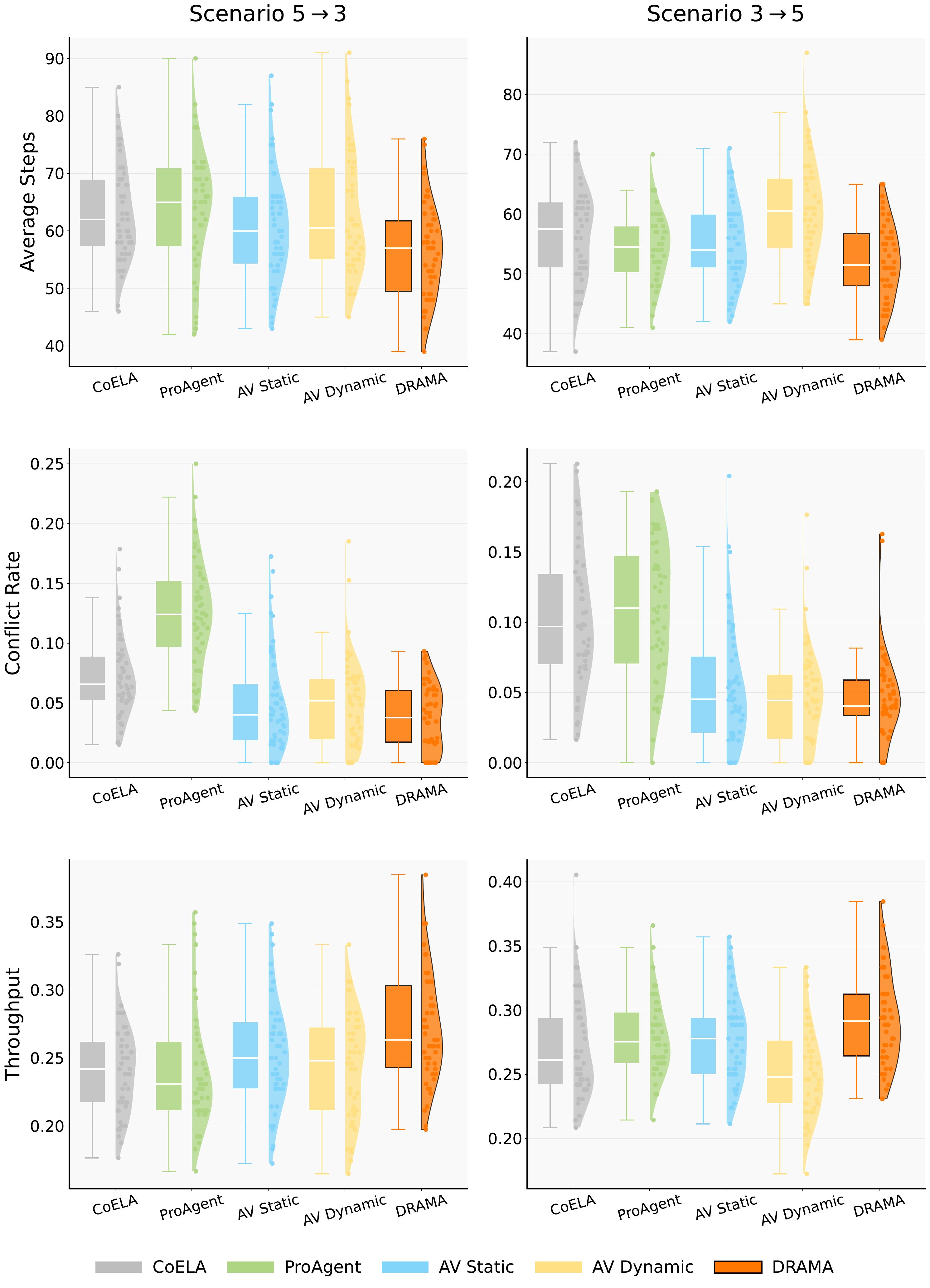}
    \caption{Distributions of Average Steps, Conflict Rate and Throughput across 50 runs for each framework on the same task.}

    \label{fig:stable}
\end{figure}

To assess the stability of the framework under challenging conditions, we selected a complex task involving a large number of required objects and conducted 50 independent runs under two dynamic scenarios: agent dropout (5$\rightarrow$3) and agent addition (3$\rightarrow$5). As illustrated in Figure~\ref{fig:stable}, DRAMA consistently outperforms all baselines, exhibiting lower medians and tighter distributions for \textit{AS} and \textit{CR}, along with higher and more stable \textit{TP} values. These results demonstrate DRAMA’s ability to sustain efficient and well-coordinated performance amid agent population fluctuations, underscoring its superior robustness and adaptability in dynamic multi-agent environments.

\subsection{Generalization of DRAMA}


To evaluate the model-agnostic generalization of DRAMA, we tested it with four diverse foundation models—GPT-4.1, GPT-4o-mini, Qwen-Max, and DeepSeek-V3.2—spanning different architectures and training paradigms. DRAMA was paired with each model and assessed under two challenging dynamic scenarios: agent dropout (5$\rightarrow$3) and agent addition (3$\rightarrow$5). As shown in Table~\ref{tab:models}, the results are highly consistent across all metrics (AS, CR, TP). These findings demonstrate that DRAMA sustains stable performance and efficiency regardless of the underlying model.

\begin{table}[t]
\centering
\caption{Performance of the DRAMA framework when integrated with different backbone foundation models. }
\renewcommand{\arraystretch}{1.2}
\setlength{\tabcolsep}{2pt}
\begin{tabular}{c|cccc}
\toprule
\textbf{Metrics} & \textbf{GPT-4.1} & \textbf{Qwen-max} & \textbf{Deepseek-v3.2} & \textbf{4o-mini} \\
\midrule

\rowcolor{gray!10}
\multicolumn{5}{c}{\textbf{Dynamic: 5 Agents to 3 Agents}} \\

AS      & 53.422    & 55.911    & 53.583  & 52.796 \\
CR      & 0.076     & 0.077     & 0.045   & 0.076 \\
TP      & 0.216     & 0.208     & 0.217   & 0.216 \\
\midrule

\rowcolor{gray!10}
\multicolumn{5}{c}{\textbf{Dynamic: 3 Agents to 5 Agents}} \\

AS      & 54.022    & 54.783    & 51.041  & 53.377 \\
CR      & 0.029     & 0.079     & 0.036   & 0.059 \\
TP      & 0.207     & 0.208     & 0.228   & 0.211 \\
\bottomrule
\end{tabular}
\label{tab:models}
\end{table}
\subsection{Ablation Study}
To further understand the contribution of each core component in DRAMA, we conduct an ablation study focusing on two key mechanisms: the Collective Intelligence Layer and the Trust Chain mechanism. The corresponding quantitative results are summarized in Table~\ref{tab:ablation_partial}, while the complete evaluation across all scenarios is reported in the supplementary. The removal of the \textit{C.I.~Layer} notably increases \textit{AS} and decreases \textit{TP}, indicating that collective spatial reasoning is essential for enhancing efficiency and task throughput. In contrast, eliminating the \textit{Trust Chain} primarily raises \textit{CR} under dynamic (4$\rightarrow$3) conditions, highlighting its pivotal role in preserving coordination and fault tolerance. Overall, these results demonstrate that DRAMA’s efficiency largely arises from the \textit{C.I.~Layer}, whereas its robustness is chiefly maintained by the \textit{Trust Chain} mechanism.

\begin{table}[t]
\centering
\caption{\textbf{Summary of Ablation Study.} 
Key results comparing the full model and its ablated variants. 
``C.I. Layer'' denotes the \textit{Collective Intelligence Layer}.}
\label{tab:ablation_partial}
\scriptsize
\resizebox{\columnwidth}{!}{%
\begin{tabular}{lccccc}
\toprule
& & \multicolumn{2}{c}{\textbf{Static}} & \multicolumn{1}{c}{\textbf{Dropout}} \\
\cmidrule(lr){3-4} \cmidrule(lr){5-5}
\textbf{Method} & \textbf{Metrics} 
& \textbf{3} & \textbf{4} & \textbf{4$\rightarrow$3} \\
\midrule
\multirow{3}{*}{DRAMA}
& AS & 59.977 & 50.242 & 57.205 \\
& CR & 0.027 & 0.062 & 0.026 \\
& TP & 0.189 & 0.217 & 0.200 \\
\midrule
\multirow{3}{*}{w/o C. I. Layer}
& AS & 65.432 & 56.674 & 61.021 \\
& CR & 0.037 & 0.054 & 0.029 \\
& TP & 0.171 & 0.205 & 0.188 \\
\midrule
\multirow{3}{*}{w/o Trust Chain}
& AS & 61.644 & 53.367 & 60.851 \\
& CR & 0.022 & 0.046 & 0.044 \\
& TP & 0.182 & 0.214 & 0.189 \\
\bottomrule
\end{tabular}%
}
\end{table}

\section{Related Work}

\textbf{Embodied Multi-Agent System}
Multi-Agent Systems (MAS) refer to computational frameworks comprising multiple autonomous agents that interact, cooperate, or compete to achieve individual or collective objectives~\cite{torreno2017cooperative}. Traditionally, MAS research has focused on developing agents capable of perceiving their environments, making decisions, and coordinating actions to address complex problems beyond the capacity of a single agent. The rapid advancement of LLMs has recently driven a paradigm shift in MAS research. LLMs now serve as powerful and adaptable backbones for agent systems~\cite{li2024survey, guo2024large}, providing strong capabilities in reasoning, planning, and communication. Recent LLM-based MAS frameworks have achieved notable success in collaborative problem-solving~\cite{qian2023chatdev, hong2023metagpt, chen2024scalable, zhang2024aflow}. The success of LLM-based multi-agent systems has expanded to embodied multi-agent systems (EMAS)~\cite{guo2024embodied, zhang2023building, chen2022cooperative, liu2022embodied, park2023generative, kaiya2023lyfe}. Through continuous interaction with their environments, EMASs acquire generalized skills for diverse tasks, advancing the pursuit of artificial general intelligence~\cite{li2025embodied}. Recent efforts increasingly focus on automating agent workflows and architectural design. Dasgupta et~al.~\cite{dasgupta2023collaborating} leverage LLMs for flexible task allocation among embodied agents. Frameworks such as CAMEL~\cite{li2023camel} and AgentVerse~\cite{chen2023agentverse} explore scalable, role-based collaboration, while MASTER~\cite{gan2025master} applies LLM-driven Monte Carlo Tree Search for adaptive agent recruitment and reward estimation. ProAgent~\cite{zhang2024proagent} emphasizes real-time belief updating and adaptive decision-making, and MaAS~\cite{zhang2025multi} introduces an \emph{agentic supernet} to enable query-conditioned architecture optimization. In parallel, Jana et~al.~\cite{jana2024online} propose online scheduling algorithms for hybrid truck–drone delivery.

\section{Conclusion}
This work presents \textbf{DRAMA}, a modular and adaptive framework for dynamic multi-agent coordination. By unifying agents and tasks under a common resource abstraction and introducing an affinity-driven, event-triggered scheduling strategy, DRAMA enables efficient, resilient, and scalable task orchestration in continuously changing environments. Extensive experiments verify its advantages in reducing execution time, mitigating conflicts, and maintaining stable throughput under agent turnover and varying task demands. These results demonstrate DRAMA’s potential as a practical framework for real-world multi-agent and embodied intelligence systems, and open promising directions for future research in decentralized, large-scale coordination and resource management.


\section*{Acknowledgment}
This work was supported in part by the National Science Foundation of China under Grants (62472375), and in part by Zhejiang Provincial Natural Science Foundation of China under Grant No. LD24F020014 and No. LD25F020002, and in part by the Zhejiang Pioneer (Jianbing) Project (2024C01032), and in part by the Ningbo Yongjiang Talent Programme(2023A-198-G), and is supported by the Zhejiang Key Laboratory Project (2024E10001).

\newpage
{
    \small
    \bibliographystyle{ieeenat_fullname}
    \bibliography{main}
}


\end{document}